
\input jnl

\def\3he{{$^3${\rm He}}}

\def\eg{{\it e.g.,\ }}
\def\ie{{\it i.e.,\ }}

\def\slD{\raise.15ex\hbox{$/$}\kern-.53em\hbox{$D$}}
\def\slA{\raise.15ex\hbox{$/$}\kern-.53em\hbox{$A$}}
\def\dsl{\raise.15ex\hbox{$/$}\kern-.57em\hbox{$\Delta$}}
\def\slp{{\raise.15ex\hbox{$/$}\kern-.57em\hbox{$\partial$}}}
\def\nsl{\raise.15ex\hbox{$/$}\kern-.57em\hbox{$\nabla$}}
\def\sla{\raise.15ex\hbox{$/$}\kern-.57em\hbox{$\rightarrow$}}
\def\slla{\raise.15ex\hbox{$/$}\kern-.57em\hbox{$\lambda$}}
\def\slb{\raise.15ex\hbox{$/$}\kern-.57em\hbox{$b$}}
\def\slr{\raise.15ex\hbox{$/$}\kern-.57em\hbox{$r$}}
\def\lnp{\raise.15ex\hbox{$/$}\kern-.57em\hbox{$p$}}
\def\lnk{\raise.15ex\hbox{$/$}\kern-.57em\hbox{$k$}}
\def\lnK{\raise.15ex\hbox{$/$}\kern-.57em\hbox{$K$}}
\def\lnq{\raise.15ex\hbox{$/$}\kern-.57em\hbox{$q$}}
\def\nna{\raise.15ex\hbox{$/$}\kern-.57em\hbox{$a$}}

\def\a{\alpha}

\def\eps{{\epsilon}}

\def\la{\lambda}

\def\bbR{{I\kern-0.3em R}}


\def\fulltriangle{{{{{{{{{\pmb{\triangle}\kern-.65em\bullet}\kern-.4em
{\raise1.2ex\hbox{.}}}
\kern-.4em{\raise1.0ex\hbox{.}}}\kern-.2em{\raise1.0ex\hbox{.}}}
\kern-.4em{\raise.1ex\hbox{.}}}\kern-.4em{\raise.2ex\hbox{.}}}
\kern-.2em{\raise.35ex\hbox{.}}}\kern.1em{\raise.2ex\hbox{.}} }}
                                                                \

\def\hexagon{{\tenpoint
\langle\kern-.1em{\raise.2cm\hbox{$\overline{\hskip.7em\relax}$}}
\kern-.7em{\lower.3ex\hbox{$\underline{\hskip.7em\relax}$}}\kern-.075em
 \rangle}}

\def\pentagon{{\tenpoint
\raise.5ex\hbox{$\widehat{\qquad}$}\kern-1.8em{\backslash
\kern-.1em{\lower.3ex\hbox{$\underline{\kern.75em}$}}\kern-.05em/} }}

\def\pmb#1{\setbox0=\hbox{$#1$}%
\kern-.025em\copy0\kern-\wd0
\kern.05em\copy0\kern-\wd0
\kern-.025em\raise.0433em\box0 }

\def\q2{{Q^2}}
\def\gtwid{\raise.3ex\hbox{$>$\kern-.75em\lower1ex\hbox{$\sim$}}}
\def\ltwid{\raise.3ex\hbox{$<$\kern-.75em\lower1ex\hbox{$\sim$}}}
\def\12{{1\over2}}

\def\part{\partial}

\def\low#1{\lower.5ex\hbox{${}_#1$}}

\def\partt{\raise.15ex\hbox{$\widetilde$}{\kern-.37em\hbox{$\partial$}}}

\def\topppageno1{\global\footline={\hfil}\global\headline
={\ifnum\pageno<\firstpageno{\hfil}\else{\hss\twelverm --\ \folio
\ --\hss}\fi}}

\def\toppageno2{\global\footline={\hfil}\global\headline
={\ifnum\pageno<\firstpageno{\hfil}\else{\rightline{\hfill\hfill
\twelverm \ \folio
\ \hss}}\fi}}

\def\boxit#1{\vbox{\hrule\hbox{\vrule\kern3pt
  \vbox{\kern3pt#1\kern3pt}\kern3pt\vrule}\hrule}}

\rightline{BROWN-HET-862}

\title Electroweak baryogenesis with electroweak strings

\author Robert H. Brandenberger$^{1),2)}$
\vskip 1cm
{\rm and}
\author Anne-Christine Davis$^{3)}$
\vskip .5in
\affil
\centerline {1) Physics Department, Brown University}
\centerline {Providence, RI 02912, USA}
\affil
\centerline {2) Institute for Theoretical Physics}
\centerline {University of California Santa Barbara, CA 93106-4030, USA}
\affil
\centerline {3) Department of Applied Mathematics}
\centerline {and Theoretical Physics \& Kings College}
\centerline {University of Cambridge, Cambridge, CB39EW, U.K.}

\abstract{
If stable electroweak strings are copiously produced
during the electroweak phase transition, they may contribute significantly
to the presently observed baryon to entropy ratio of the universe. This
analysis establishes the feasibility of implementing an electroweak
baryogenesis
scenario without a first order phase transition.
}

\endtopmatter

\head{1. Introduction}

Recently there has been a revival of interest in studies of the electroweak
phase transition. Two of the main reasons are the discovery that a nonvanishing
baryon to entropy ratio can be generated at the electroweak scale,\refto{1-4)}
and the rediscovery of solitonic solutions (electroweak
strings\refto{5)}) in the standard electroweak theory (which were initially
discussed by Nambu\refto{6)}).

Electroweak baryogenesis has immense relevance for cosmology. The goal is to
explain the observed nonvanishing baryon to entropy ratio $n_B/s$, $n_B$ being
the net baryon number density, $s$ the entropy density. As realized by
Sakharov,\refto{7)} the three necessary ingredients to be able to generate a
nonvanishing $n_B/s$ are the existence of baryon number violating interactions,
$CP$ violation and the presence of out of equilibrium processes.

Until recently, the canonical implementation of baryogenesis has been in the
context of grand unified models. The out of equilibrium decay of superheavy
gauge and Higgs particles\refto{8)} was the main physical mechanism operating,
making use of the explicit $n_B$ violation in the Lagrangian. A recently
discovered variant\refto{9)} of GUT baryogenesis is based on the collapse of
topological defects---in particular cosmic strings.

Of late, however, some problems have arisen for this scenario. Most
importantly, it was realized that electroweak anomaly
effects at temperatures above the critical temperature of the phase
transition
 effectively erase\refto{10)} any primordial baryon symmetry if $B-L=0$ ($B$
and $L$ being baryon and lepton numbers respectively). Hence it is particularly
important to study the possibility of regenerating a nonvanishing $n_B/s$ ratio
after the electroweak phase transition.

The mechanisms suggested so far for electroweak baryogenesis all rely on
having a
first order phase transition. The resulting bubble walls were required in order
to obtain  a region of unsuppressed baryon number violation occurring out of
thermal equilibrium. Our work is based on the observation that topological
defects forming in a second order phase transition may play a similar role to
the bubble walls. We propose a specific mechanism in which electroweak strings
are responsible for baryogenesis.

Electroweak strings\refto{5)} are nontopological solitons which arise in the
standard electroweak theory (and extensions thereof). They are essentially
Nielsen-Olesen\refto{11)} strings of $U(1)_Z$ embedded in the
 $SU(2)\times U(1)$
theory ($U(1)_Z$ is the Abelian subgroup which is broken during the electroweak
phase transition). For certain ranges of the parameters of the standard model,
electroweak strings are energetically stable.\refto{12)} Like semilocal
strings,\refto{13)} electroweak strings are not topologically stable.

If, however, we are in a region of parameter space in which electroweak strings
are stable, a network of such strings will form during the electroweak phase
transition---even if it is second order. Inside the strings, anomalous baryon
number violating processes are unsuppressed. If the strings move, the out of
thermal equilibrium condition will be satisfied. Finally, the standard model
contains $CP$ violation. Hence all of Sakharov's criteria are satisfied. As we
shall demonstrate, it is in fact possible to generate a substantial $n_B/s$
using electroweak strings.

In the following we shall first briefly review the proposed electroweak
baryogenesis scenarios (section 2) and electroweak strings (section 3). In
section 3 we propose two specific mechanisms of baryogenesis based on
electroweak strings. We conclude with a discussion of the results. Units in
which $c=\hbar=k_B=1$ are used throughout.

\head{2. Electroweak Baryogenesis}

Although at a classical level, the electroweak theory conserves baryon
number, there are baryon number violating quantum effects due to the
electroweak anomaly.\refto{14)} The standard model also contains $CP$ violating
effects. Hence, provided some of the $CP$ and baryon number violating processes
occur out thermal equilibrium, it is possible to generate a nonvanishing
$n_B/s$ ratio after the electroweak phase transition.

Following some initial suggestions,\refto{1)} interesting specific mechanisms
were proposed by Turok and Zadrozny,\refto{2)} and by Cohen, Kaplan, and
Nelson.\refto{3,4)} These mechanisms\refto{2-4)} rely on having a first order
electroweak phase transition producing bubbles of broken symmetry phase
expanding into the unbroken phase. The bubble walls form the region where the
out-of-equilibrium, $CP$-violating effects take place and where the net baryon
number is generated.

The Turok and Zadrozny mechanism makes use of nontrivial winding number
(``local texture") configurations which are in equilibrium in the unbroken
phase. Unwinding of such a configuration involves a change in Chern-Simons (and
hence baryon) number. Inside the bubble wall the change in baryon number
has a preferential direction
(because of $CP$ violation in the Higgs
sector), and this leads to a net baryon asymmetry.

Cohen, Kaplan, and Nelson have proposed two mechanisms for electroweak
baryogenesis. In the first,\refto{3)} a detailed thermodynamic  argument is
presented showing that an effective chemical potential for baryon number is
generated by the $CP$ violating effects inside the bubble wall. The second
mechanism\refto{4)} is based on fermions incident from the unbroken phase
scattering off the bubble walls. $CP$ violation in the walls leads to an
asymmetry in particle-antiparticle scattering, and hence to a lepton excess in
the unbroken phase which is---via equilibration---converted into a baryon
asymmetry. The baryon asymmetry does not change as the bubble wall passes by.

The three mechanisms discussed above are all inequivalent.\refto{15)} The
maximal baryon to entropy ratio $n_B/s$ which can be obtained is of the
order\refto{4)}
$$
{n_B\over s}\sim \eps\a_w^4 \leqno(1)
$$
where $\eps$ measures the strength of $CP$ violation (in the two Higgs doublet
model,\refto{2-4)} $\eps$ can be as large as 1) and $\a_w=g_w^2/4\pi$, $g_w$
being the weak coupling constant. Thus, it is not too hard to generate the
observed baryon to entropy ratio at the electroweak scale.

However, it is unclear\refto{16)} whether the electroweak phase transition is
sufficiently strongly first order for the above mechanisms to work. It is
therefore interesting to explore the possibility of generating a nonvanishing
$n_B/s$ below the electroweak symmetry breaking scale assuming that the
transition proceeds without the formation of bubbles. We shall propose a
concrete mechanism which makes use of electroweak strings.

\head{3. Electroweak Strings}

Electroweak strings\refto{5,6)} are essentially an embedding of the
Nielsen-Olesen $U(1)$ string\refto{11)} in the standard electroweak theory.
They are solutions of the field equations for all electroweak parameters, but
are stable only for a narrow range of these parameters.\refto{12)}

The Lagrangian for the bosonic part of the Weinberg-Salam model contains
$SU(2)$ gauge fields
$W^i$, $i=1,\dots,3$, a $U(1)$ gauge field $B$ and a complex scalar doublet
$\phi$. The vortex solution which extremizes the action is
$$
\phi=f_{NO}(r)e^{im\theta} \pmatrix{0\cr1},\quad \vec Z=\vec A_{NO} \leqno(2)
$$
and $\vec A=0=\vec W^a (a=1,2)$. Here, $r$ and $\theta$ are polar coordinates
in the plane perpendicular to the string,
 $\vec A_{NO}$ is the Nielsen-Olesen
gauge field,
$$
\phi_{NO}=f_{NO}(r)e^{im\theta} \leqno(3)
$$
is the Nielsen-Olesen Higgs field and
$$
\vec Z=\cos \theta_w \vec W^3-\sin\theta_w \vec B,\ \ \vec A=\sin\theta_w \vec
W ^3+\cos\theta_w\vec B. \leqno(4)
$$
$\theta_w$ is the weak mixing angle, and $f_{NO}(r)$ is a function which
approaches
the symmetry breaking scale $\eta$ at large $r$ and vanishes for $r=0$.
It has been shown\refto{12)} that these vortex solutions are stable for
$\sin^2\theta_w\simeq 1$, in which case the electroweak string is essentially
the $U(1)_B$ Nielsen-Olesen vortex.

In order to obtain a sufficiently large baryon to entropy ratio, the standard
electroweak model must be extended by adding new terms in the Lagrangian which
contain explicit $CP$ violation. An often used prototype theory is the two
Higgs model.\refto{2-4)}

The above construction of nontopological vortex solutions in theories which do
not satisfy the topological criterion for strings is not specific to the
minimal standard model. Thus, we expect electroweak strings to exist also in
extensions of the Weinberg-Salam model (This has recently been demonstrated
in the two Higgs model \refto{17)}). It is possible that these strings could
be stable even for experimentally allowed values for the model parameters. In
the following we shall assume that electroweak strings exist and are stable.

In models admitting stable electroweak strings, a network of such strings will
form during the electroweak phase transition. If we consider a theory with
Higgs potential
$$
V(\phi)=\la(\phi^+\phi-\eta^2/2)^2, \leqno(5)
$$
then the initial correlation length (mean separation of strings) will
be\refto{18)}
$$
\xi(t_G)\simeq\la^{-1}\eta^{-1}, \leqno(6)
$$
where $t_G$ is the time corresponding to the Ginsburg temperature of the phase
transition.

The initial network of electroweak strings will be quite different from that of
cosmic strings, the reason being that electroweak strings can end on local
monopole and antimonopole configurations. From thermodynamic
considerations,\refto{19)} we expect most of the strings to be short, \ie of
length $l\simeq\xi(t_G)$, since this maximizes the entropy of the network for
fixed energy.

After the phase transition, the vortices will contract along their axes and
decay after a time interval
$$
\Delta t_S\simeq{1\over v}(\la \eta)^{-1} \leqno(7)
$$
where $v$ is the velocity of contraction (expected to be $\simeq1$). In the
following, we shall demonstrate that the string contractions will produce a net
baryon symmetry.

\head{4. The Baryogenesis Mechanism}

We shall consider an extension of the standard electroweak theory in which
there is additional $CP$ violation in the Higgs sector. An example is the two
Higgs model used in Refs.~2-4.
We assume that electroweak strings can be embedded in this model \refto{17)},
and we choose
the values of the parameters in the Lagrangian for which these strings are
stable. Furthermore, the phase transition is taken to be second order.

A key issue is the formation probability of electroweak strings. In the
following, we make the rather optimistic assumption that both the mean length
and average separation of electroweak strings at $t_G$ will equal the
correlation length $\xi(t_G)$. For topological defects, this result follows
from the Kibble mechanism \refto{18)}. When applied to electroweak strings, the
Kibble mechanism implies that the vortex fields $\phi$ and $Z$ have the
correlation length $\xi(t_G)$. However, to form an electroweak string, the
other fields must be sufficiently small such that the configuration relaxes to
the exact electroweak string configuration. Obviously, the restriction this
imposes (and the consequent increase in the mean separation of electroweak
strings) is parameter dependent - the more stable the strings, the smaller
the increase in the mean separation.
Pieces of string are bounded by
monopole-antimonopole pairs. Energetic arguments tell us that the string will
shrink. We now argue that the moving string ends will have the same effects on
baryogenesis as the expanding bubble walls in Refs.~2\&3.

We can phrase our argument either in terms of the language of \Ref2 or of
\Ref3. The phase of the extra $CP$ violation is nonvanishing in the region in
which the Higgs fields $\phi$ are changing in magnitude, \ie at the edge of the
string. Since $|\phi|$ increases in magnitude, $CP$ violation has a definite
sign. Hence, in the language of \Ref3, a chemical potential with definite sign
for baryon number is induced at the tips of the string (where $|\phi|$ is
increasing). This chemical potential induces a nonvanishing baryon number.

In the language of \Ref2, the $CP$ violation with definite sign at the tips of
the string leads to preferential decay of local texture configurations with a
definite net change in Chern-Simons (\ie baryon) number.

Let us now estimate the magnitude of this effect. The rate of baryon number
violating events inside the string (in the unbroken phase) is
$$
\Gamma_B\sim\a_w^4T^4.\leqno(8)
$$
The volume in which $CP$ violation is effective changes at a rate ($g$ is the
gauge coupling constant)
$$
{dV\over dt}=g^2w^2V,\leqno(9)
$$
where $w\simeq\la^{-1/2}\eta^{-1}$ is the width of the string and $v$ is its
contraction velocity. The
factor $g$ comes from the observation that baryon number violating processes
are unsuppressed only if $|\phi|<g\eta$.\refto{20)}
The rate of baryon number generation per string is
$$
{dN_B\over dt}\sim w^2v \Gamma_B\eps\Delta t_c,\leqno(10)
$$
where $\eps$ is a dimensionless constant measuring the strength of $CP$
violation and
$$
\Delta t_c={gw\over v} {1\over \gamma(v)} \leqno(11)
$$
is the time a fixed point in space is in the transition region. Here,
$\gamma(v)$
is the usual relativistic $\gamma$ factor. Since there is one string per
correlation volume $\xi(t_G)^3$, the resulting rate of increase in the
baryon number density $n_B$ is
$$
{dn_B\over dt}\sim \la^{-3/2} \eta^{-3} g^3 \a_w^4 T^4 {1\over
\gamma(v)}
\eps \xi (t_G)^{-3}.\leqno(12)
$$
The net baryon number density is obtained by integrating (12) from $t_G$, the
time corresponding to the Ginsburg temperature, and $t_G+\Delta t_S$ (see (7)).
The result is
$$
n_B \sim {\la\over {v\gamma(v)}} g^3 \a_w^4 T^3_G\eps.\leqno(13)
$$

Our result (13) must be compared to the entropy density at $t_G$:
$$
s(t_G)= {\pi^2\over 45} g^* T^3_G,\leqno(14)
$$
where $g^*$ is the number of relativistic spin degrees of freedom. From (13)
and (14) we obtain
$$
{n_B\over s}\sim
{45\over\pi^2g^*}{\la\over {\gamma(v)v}}\eps g^3\a_w^4.\leqno(15)
$$
For $\la\sim v\sim1$ and $\eps\sim 1$, the ratio obtained is only slightly
smaller than the observational value.

In order for our mechanism to work, the core radius of the string ($\vert \phi
\vert < g \eta$) must be large enough to contain the nonperturbative
configurations which mediate baryon number violating processes. This leads to
the condition $\la < g^4$, i.e. small Higgs mass. In addition, the sphaleron
must be sufficiently heavy such that sphaleron transitions in the broken
symmetry phase are suppressed for $T = T_G$. For small values of $\la$, this
condition will automatically be satisfied. Finally, the model parameters must
be such that the phase transition is of second order. In the standard
electroweak theory, this condition is incompatible with $\la \ll g^4$. In any
extended electroweak theory, the consistency of the above conditions must be
satisfied in order for our baryogenesis mechanism to be effective.

\head{5. Discussion}

We have presented a counterexample to the ``folk theorem" stating that
electroweak baryogenesis requires a first order electroweak phase transition.
We propose a mechanism in which finite length electroweak strings during their
contraction generate a nonvanishing net baryon number. The strings play
a similar role to the expanding bubble walls in a first order phase transition:
they provide out of equilibrium processes, and also a region where $CP$
violation occurs.

The mechanism presented here requires stable electroweak strings and an extra
source of $CP$ violation (which is present in the two Higgs models used in
Refs.~2-4). Based on the stability analysis of electroweak strings in the
standard model,\refto{12)} it is unlikely that these strings will be stable for
experimentally allowed values of the parameters in the Lagrangian.

Note that translational or rotational motion of the strings will not generate
any net asymmetry since in this case the absolute value of the Higgs fields
will increase at some points in space (those leaving the string) and decrease
at others (those entering the string). Hence, the contributions to the net
baryon number should cancel out. In our mechanism, it is essential that below
$T_G$ there is a distinguished direction to the evolution of $|\phi|$, and
hence a distinguished sign for the chemical potential for baryon number.

We hope that this work will point toward more realistic mechanisms of
electroweak baryogenesis using second order phase transitions.

\head{Acknowledgements}

For interesting discussions we are grateful to
 R.~Holman and L.~McLerran.
 This
work was supported in part (at Brown) by DOE grant DE-AC02-76ER03130 Task A, by
an Alfred P. Sloan Fellowship (R.B.), and by an NSF-SERC Collaborative Research
Award NSF-INT-9022895 and SERC GR/G37149. One of us (R.B.) acknowledges the
hospitality of the ITP where this work was completed with support from
NSF Grant No. PHY89-04035.

\references

\refis{1} M. Shaposhnikov, \journal JETP Lett., 44, 465, 1986; M. Shaposhnikov,
\np B287, 757, 1987; L. McLerran, \prl 62, 1075, 1989.

\refis{2} N. Turok and J. Zadrozny, \prl 65, 2331, 1990; N. Turok and J.
Zadrozny, \np B358, 471, 1991; L. McLerran, M. Shaposhnikov, N. Turok and M.
Voloshin, \pl 256B, 451, 1991.

\refis{3} A. Cohen, D. Kaplan and A. Nelson, \journal Phys. Lett.,
 B263, 86, 1991.

\refis{4} A. Nelson, D. Kaplan and A. Cohen, UCSD Preprint UCSD/PTH 91-20
(1991).

\refis{5} T. Vachaspati, \prl 68, 1977, 1992.

\refis{6} Y. Nambu, \np B130, 505, 1977.

\refis{7} A. Sakharov, \journal Pisma ZhETF, 5, 32, 1967.

\refis{8} S. Dimopoulos and L. Susskind, \pr D18, 4500, 1978; M. Yoshimura
\prl 41, 281, 1978; A. Ignatiev, N. Krasnikov, V. Kuzmin, and A. Tavkhelidze,
\journal Phys. Lett., B76, 436, 1978; S. Weinberg, \prl 42, 850, 1977; D.
Toussaint, S. Trieman, F. Wilczek, and A. Zee, \pr D19, 1036, 1979.

\refis{9} R. Brandenberger, A.-C. Davis, and M. Hindmarsh, \journal Phys.
Lett., B263, 239, 1991.

\refis{10} V. Kuzmin, V. Rubakov, and M. Shaposhnikov, \journal Phys. Lett.,
B155, 36, 1985; P. Arnold and L. McLerran, \prd 36, 581, 1987.

\refis{11} H. Nielsen and P. Olesen, \np B61, 45, 1973.

\refis{12} L. Perivolaropoulos and T. Vachaspati, Santa Barbara preprint
NSF-ITP-92-31 (1992).

\refis{13} T. Vachaspati and A. Ach\'ucarro, \prd 44, 3067, 1991.

\refis{14} G. 't Hooft, \prl 37, 8, 1976.

\refis{15} R. Brandenberger and A.-C. Davis, `A note on electroweak
baryogenesis', Brown preprint BROWN-HET-827 (1992).

\refis{16} see \eg  M. Dine, R. Leigh, P. Huet, A. Linde, and D. Linde, SLAC
preprint SLAC-PUB-5741 (1992).

\refis{17} A.-C. Davis and U. Wiedemann, DAMTP preprint (1992).

\refis{18} T.W.B. Kibble, \journal J. Phys., A9, 1387, 1976.

\refis{19} D. Mitchell and N. Turok, \prl 58, 1577, 1987; M. Sakellariadou and
A. Vilenkin, \prd 42, 349, 1990.

\refis{20} M. Dine, P. Huet, and R. Singleton, Santa Cruz preprint SCIPP
91/08 (1991).

\endreferences

\endit